\documentclass{aa}

\usepackage{comment}

\usepackage{graphicx}
\usepackage{txfonts}
\usepackage{lipsum}
\usepackage{subcaption}       
\usepackage{lscape}            
\usepackage{placeins}           
\usepackage{gensymb}

\begin{document}

   \title{First Sardinia Radio Telescope detection of the Sunyaev-Zel’dovich effect at 18.6 GHz}

   \author{S. Cocchi\inst{1,2,3} \fnmsep\thanks{E-mail: silvia.cocchi@inaf.it}
        \and F. Loi\inst{2}
        \and M. Murgia\inst{2}
        \and P. Marchegiani\inst{2}
        \and V. Vacca\inst{2}
        \and F. Govoni\inst{2}
        \and F. Gandossi\inst{1,2,3}
        \and G. Rodighiero \inst{4}
        }

   \institute{Dipartimento di Fisica e Astronomia, Università di Bologna, via P. Gobetti 93/2, 40129, Bologna, Italy
    \and INAF-Osservatorio Astronomico di Cagliari, Via della Scienza 5, I-09047 Selargius (CA), Italy
      \and INAF, Istituto di Radio Astronomia, Via Gobetti 101, 40129 Bologna, Italy
   \and  Dipartimento di Fisica e Astronomia, Università degli Studi di Padova, Vicolo dell’Osservatorio 3, I-35122 Padova, Italy
            }

   \date{Received January 15, 2026 / Accepted February 21, 2026}

  \abstract
   {Galaxy clusters imprint a distinctive signature on the cosmic microwave background through the thermal Sunyaev–Zel’dovich (SZ) effect, which enables to study the intracluster plasma distribution and makes them powerful cosmological probes.
   We present the first Sardinia Radio Telescope (SRT) detection of the SZ effect in the galaxy cluster MACS~J1752+4440 at 18.6 GHz, with a resolution of 0.9\arcmin.
   We detected a decrement in brightness toward the cluster centre, which we attributed to the thermal SZ effect. We modelled the signal using a spherically symmetric $\beta$ model for the electron density distribution and we employed a Bayesian retrieval to estimate the core radius, central electron density, and $\beta$ parameter of the cluster.
   We found values consistent with expectations for a galaxy cluster of the mass of MACS~J1752+4440: a core radius of ($160 \pm 30$) kpc, a central electron density of ($2.5^{+0.7}_{-0.5} \cdot 10^{-3}$) cm$^{-3}$, and $\beta$=$0.6\pm0.1$. The mean Compton-$y$ parameter within a radius of 3.5$\arcmin$ is $(2.6 \pm 0.3) \cdot 10^{-5}$,  higher than the value reported by Planck, which is coherent considering the different resolution of the instruments and the modelling adopted.
   This work demonstrates the potential of the SRT to detect the onset of the SZ decrement at low frequencies, providing higher angular resolution than current all-sky surveys and enabling an improved reconstruction of the SZ decrement profile and the plasma distribution in the intracluster medium.}


   \maketitle

\section{Introduction}
Clusters of galaxies, as the largest gravitationally bound systems in the Universe, are investigated at all frequencies, from the radio to the optical to the X-ray. In particular, at millimeter wavelengths, they imprint signatures on the Cosmic Microwave Background (CMB) through the Sunyaev–Zel’dovich (SZ) effect \citep{sunyaev1972}. The SZ effect results from the inverse Compton scattering of CMB photons by thermal electrons in the intracluster medium (ICM). By shifting photons to higher frequencies, this process generates a characteristic spectral distortion: a decrement in CMB intensity at low frequencies that transitions to an increment above $\sim$220 GHz.
This effect can be separated into two components: the kinetic SZ (kSZ) effect \citep[e.g.,][]{1980MNRAS.190..413S}, due to the cluster’s line-of-sight motion, and the thermal SZ (tSZ) effect, which traces the thermal pressure and dynamics of the hot ICM. The tSZ effect has been widely used to detect and characterise galaxy clusters, with large surveys yielding catalogues of $\sim10^2$–$10^3$ systems. It  offers a nearly mass-limited census of the cluster population at high redshift (where the abundance is sensitive to cosmological parameters), while avoiding X-ray selection biases that favour nearby, dynamically relaxed systems (e.g. \citealt{planck2016}). The Planck Collaboration selected through the SZ effect over 1600 systems between 100 and 857 GHz \citep{planck2020A&A...641A...1P}. The South Pole Telescope (SPT; \citealt{reichardt2013}) survey delivered a catalogue of over 1000 SZ-selected clusters  \citep{bleem2020ApJS..247...25B}, while the Atacama Cosmology Telescope (ACT) survey reported over 10000 SZ-detected clusters at 90-220 GHz \citep{2026OJAp....955863A}. However, these surveys have relatively poor angular resolutions: 5$\arcmin$ for Planck,
and about 1$\arcmin$ for SPT and ACT. A higher resolution imaging of the tSZ effect was obtained using, for instance, MUSTANG-2, a 223-feedhorn bolometer camera mounted on the 100-meter Green Bank Telescope (GBT) \citep{dicker2014JLTP..176..808D}. Operating at 90 GHz with $\sim9\arcsec$ resolution and a $4.2\arcmin$ field of view, MUSTANG-2 resolves ICM substructures such as shocks and pressure discontinuities that are inaccessible to survey-class experiments \citep{romero2020ApJ...891...90R}. NIKA2 (\citealt{adam2018}) provides $17.6\arcsec$ resolution at 150 GHz, enabling the detection of low-mass clusters out to intermediate and high redshifts \citep{cherouvrier2025}. 
New  Kinetic Inductance Detector (KID) based bolometric cameras are further advancing high-resolution millimetre observations: MISTRAL, installed at the 64-m Sardinia Radio Telescope (SRT), operates at 90 GHz with a $4\arcmin$ field of view and $\sim12\arcsec$ resolution, optimised for SZ and diffuse emission studies (Battistelli et al. in prep., Murgia et al in prep.); TolTEC on the 50-meter Large Millimeter Telescope (LMT) provides simultaneous, polarisation-sensitive imaging across 143, 214, and 273 GHz, enabling precise separation of the thermal SZ effect from dusty star-forming galaxies with resolution of $5\arcsec$ \citep{bryan2018}.\\
\noindent
In this Letter, we report the first SRT \citep{bolli2015,prandoni2017} detection of the SZ effect in a galaxy cluster at 18.6~GHz. The result presented here is a serendipitous discovery in the context of a project aimed at investigating the properties of the double radio relic system in the galaxy cluster MACS J1725+4440 \citep[z=0.366, ][]{edge2003}. At these frequencies, diffuse radio sources such as relics are affected by contamination from this effect, so properly characterizing it is essential for obtaining accurate flux density measurements \citep{basu2016}. MACS~J1752+4440, centred at $RA_{J2000}$ = $17^{\mathrm{h}}52^{\mathrm{m}}01.5^{\mathrm{s}}$  and
$DEC_{J2000}$ = $+44^{\circ}40'46''$ according to the ROSAT All Sky Survey \citep[RASS;][]{voges1999}, was discovered in the Massive Cluster Survey \citep{ebeling2001} and later identified as a radio bright cluster by \citet{edge2003} through a joint analysis of the
Westerbork (WSRT) Northern Sky Survey \citep{rengelink1997} and RASS Bright Source Catalog.
Radio studies with the WSRT \citep{vanweeren2012} and Giant Microwave Radio Telescope (GMRT) \citep{bonafede2012} revealed it to be a merging double relic cluster. The SZ-derived mass\footnote{M$_{500}$ is defined as the mass enclosed in R$_{500}$, where the density is 500 times the critical density of the Universe.} reported by the \citet{planck2016}  $M_{500} = (6.7^{+0.4}_{-0.5}) \cdot 10^{14}M_{\odot}$ is consistent within 1$\sigma$ with the total mass $M_{500} = (1.47^{+0.33}_{-0.38}) \cdot 10^{15} M_{\odot}$ derived from Subaru and XMM–Newton data, according to \citet{finner2021}.\\
The contents of this Letter are organised as follows.  Sect. \ref{data reduction} reports the details of observations and data reduction pipeline. The total intensity results are presented in Sect. \ref{results}, highlighting the detection of the SZ decrement in the centre of the cluster. Our modelling process is presented in Sect. \ref{modeling}.  Finally, the summary and conclusions are
given in Sect. \ref{conclusions}. Throughout this work we assume a flat $\Lambda$CDM cosmology with $H_{0} = 70\ \mathrm{km\ s^{-1}\ Mpc^{-1}}$, $\Omega_{\mathrm{m}} = 0.3$, and $\Omega_{\Lambda} = 0.7$. At the redshift of MACS J1752+4440, 1\arcmin corresponds to $\sim$305 kpc.

\section{Observations and data reduction} \label{data reduction} 
With the SRT, we observed an area of 10$\arcmin$$\times$10$\arcmin$ centred on MACS J1752.0+4440 centre with the seven-feed K-band receiver, with a bandwidth of 1200 MHz centred at 18.6 GHz (project
code 18-23, P.I. Francesca Loi). The observations were carried out between November 2024 and January 2025, with the SARDARA backend \citep[SArdinia Roach2-based Digital Architecture for Radio Astronomy,][]{melis2018}.  
We performed on-the-fly \citep[OTF;][]{magnum2007} mapping in the equatorial frame, alternating the RA and DEC directions. The use of orthogonal maps enables the identification and mitigation of scanning correlated noise through de-striping techniques \citep[as described in Appendix A1 of][]{murgia2016}. The telescope scanning speed was set to 2$\arcmin$/sec and the scans were separated by 0.25$\arcmin$ to properly sample the SRT beam with at least 4 pixels. Data were taken on 15 different days, with observations lasting 5.5 hours each (80\%  of them on-source) for a total of 82.5 hours.\\
The data reduction was carried out with the proprietary Single-dish Spectral-polarimetry Software \citep[SCUBE;][]{murgia2016}, following a procedure similar to those previously described by \citet{loi2020} and \citet{bianchi2022}. Spectral channels affected by backend issues were first flagged, followed by an automatic removal of outliers and radio frequency interferences (RFI). Observations of the cold sky with and without an injected signal were used to correct gain and delay differences between right- and left-hand polarisations. Atmospheric contributions were removed using skydip scans to estimate atmospheric opacity and system temperature \citep{buffa2016}. Since the data were collected in 15 separate observing sessions, any residual atmospheric noise is expected to behave randomly across the field of view. Baselines were subtracted using the first and last 10\% of each subscan. Bandpass calibration was performed using 3C286 to correct the frequency-dependent response of the seven K-band feeds \citep{orfei2010}. 
After applying all the previous terms, the calibrators data were then gridded with a scale of 15 $\arcsec$ per pixel. Then a flux density scale calibration was performed with 3C286. For each of the seven feeds and for both left- and right-hand circular polarisations (LCP and RCP), a 2D Gaussian was fitted to the calibrator maps using a combination of random search and gradient descent. The conversion from counts to Jansky was applied using the flux-density model from \citet{perley2013}. The flux calibration was applied to 3C295, which was used to assess the calibration accuracy  by comparing the results with the theoretical values at 17.975 GHz. A 2D Gaussian was fitted to each calibrated image and the relative errors, computed as the difference between measured and expected peak flux, then normalised by the latter, were evaluated across all feeds, polarisations, and scan pairs. The distribution of these errors was used to identify new outliers arising after flux calibration, likely due to residual RFI, and an additional manual flagging step was applied, ensuring that the remaining calibration error was kept below 10\%.
After applying the calibration solutions to the galaxy cluster data, a baseline subtraction was applied to individual subscans using second-order polynomials, excluding masked regions to avoid bright-source contamination; the mask was constructed from the 3$\sigma_{NVSS}$  contours at 1.4 GHz (see Figure \ref{total_intensity}). Masking and polynomial subtraction prevent a potential negative bias at the map centre. Baseline fitting was iteratively refined with 5$\sigma$ clipping. Gridded maps were created with a resolution of 128$\times$128 pixels and a pixel scale of 15\arcsec. We combined the calibrated LCP and RCP data to form separate Stokes I images for the RA and Dec scans using weighted stacking \citep{murgiafatigoni2024ApJS..272...10M}. The two final RA and Dec maps were then merged using a wavelet-based method \citep{murgia2016} to enhance the signal-to-noise ratio (S/N) and destripe the residual scanning noise.

\section{Total intensity results and decrement detection} \label{results}

   \begin{figure}[h]
   \centering
   \includegraphics[width=\hsize]{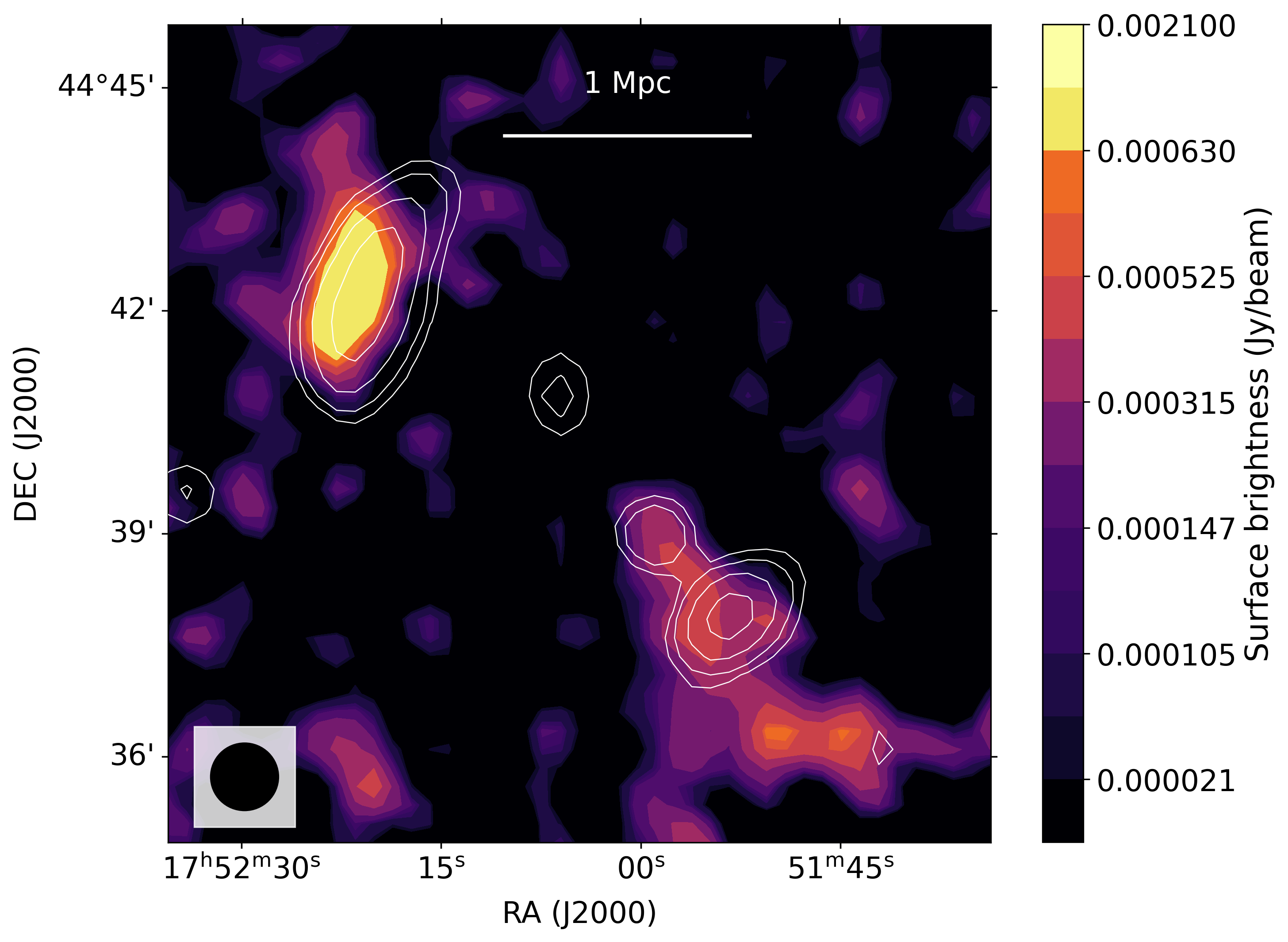}
   \captionsetup{font=footnotesize, skip=0pt}
      \caption{ SRT total intensity image at 18.6 GHz with beam of 0.9\arcmin, pixel size of 15\arcsec. Contours at 2, 3, 5, 10, and 20 $\sigma_{NVSS}$ from the NVSS image \citep{condon1998} at 1.4 GHz, with $\sigma_{NVSS}$=0.4 mJy/beam.}
         \label{total_intensity}
   \end{figure}

\begin{figure*}
\centering
                \begin{minipage}{0.46\textwidth}
                        \centering
                        \includegraphics[width=\linewidth]{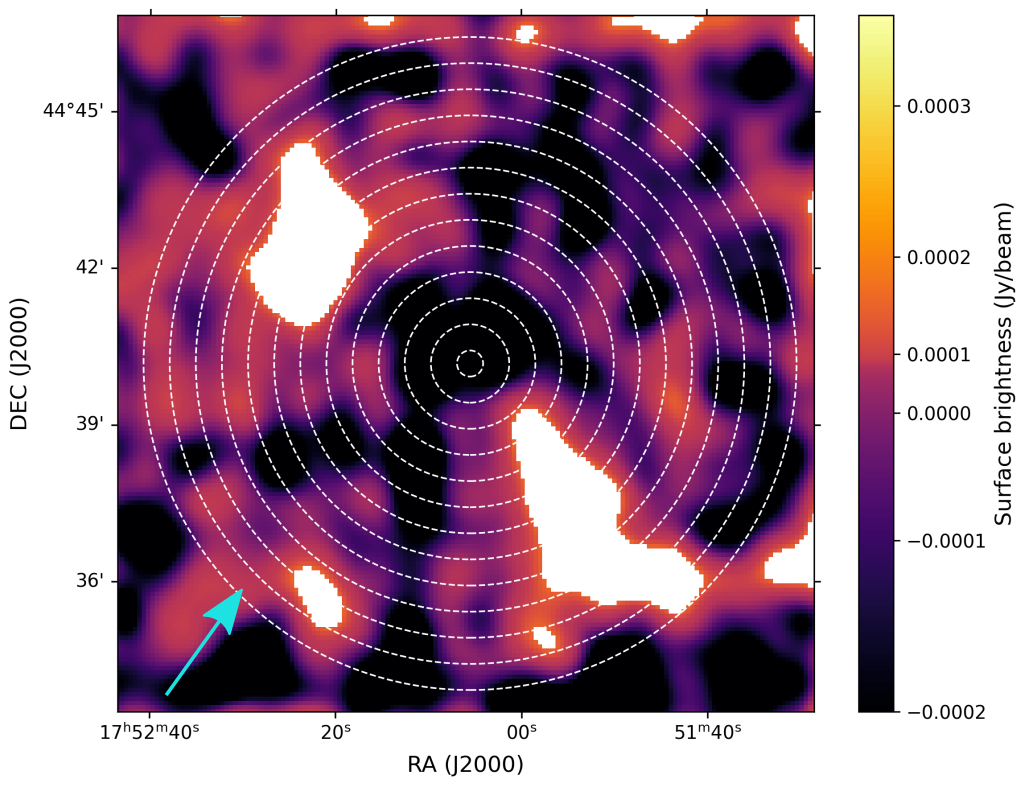}
                        
                \end{minipage}
                \hspace{0.05\textwidth} 
                \begin{minipage}{0.46\textwidth}
                        \centering
                        \includegraphics[width=\linewidth]{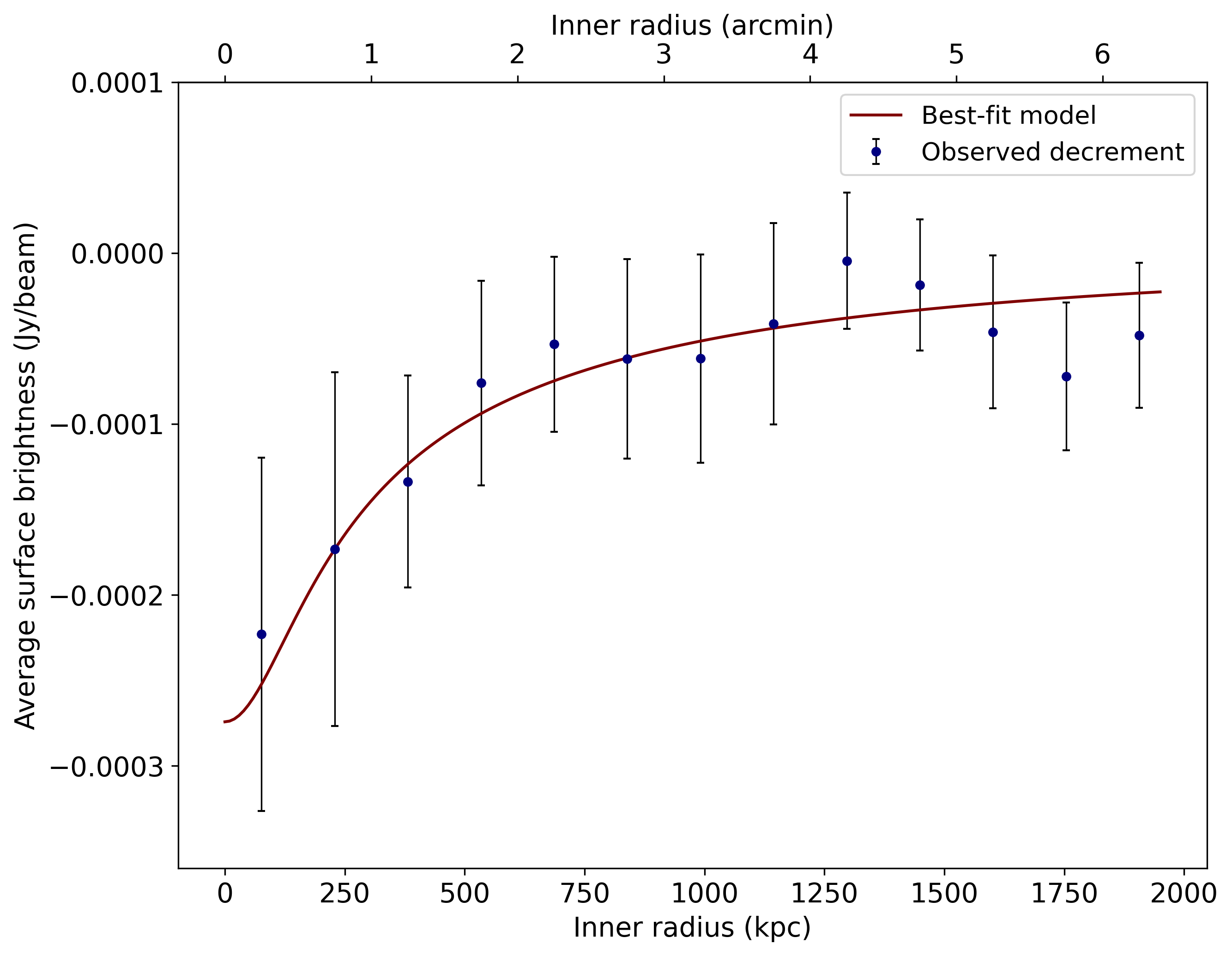}
                        
                \end{minipage} 
                \captionsetup{font=footnotesize, skip=0pt}
                \caption{\textit{Left:} Map of the observed surface brightness decrement in the cluster centre from SRT data at 18.6 GHz. The profile reported on the right was obtained by blanking the radio sources and tracing annuli around the centre of the cluster, at intervals of a half-beam radius from one another. The blue arrow marks a localised surface-brightness excess in the south-eastern region. \textit{Right: }  Observed average surface brightness on annuli around the centre of the cluster, in blue. The continuous line shows the fit of the SZ decrement referred to the models from Eqs. \ref{eq decrement} and \ref{electrondensity}. }
                \label{sz images}
    \end{figure*}

\noindent
Figure \ref{total_intensity} shows the resulting 18.6 GHz SRT image obtained by averaging the data between 18 GHz and 19.2 GHz. The noise is $\sigma$=0.2 mJy/beam, the beam size is 0.9$\arcmin$.

\noindent
The image captures the emission from the northern relic at 3$\sigma$; however, it cannot separate, at 2$\sigma$, the emission of the southern radio relic and of the source just northeast of it, which has been assumed  to be a radio galaxy in previous
works \citep[see][]{bonafede2012, vanweeren2012}. The SRT detects emission from the northern relic centred in a different region with respect to the NVSS, likely because at 18.6 GHz, the SRT is more sensitive to the outer edge of the relic, closest to the shock front where particle acceleration occurs and the spectrum is relatively flat. Toward the cluster centre, radiative losses suppress the population of electrons emitting at such high frequencies, whereas electrons emitting at 1.4 GHz are still present.
The left panel of Figure \ref{sz images} allows us to visualise an evident surface brightness decrement at the cluster centre, which we attribute to the tSZ effect. This image was produced by applying a Gaussian smoothing ($\sigma_{smooth}$=2 pixels) to the data, after masking out the emission from the radio sources from SRT data at 1$\sigma$, and subsequently upscaling the image by a factor of three using cubic interpolation to enhance visualisation.
The radial surface-brightness profile, measured in azimuthal segments, is largely circularly symmetric, except for a bump between $\sim$1200 and 1600 kpc (Figure \ref{sz images}), corresponding to a localised excess in the southeastern region of the cluster (cyan arrow), which should be taken into account for future observations of the cluster.
The profile of the detected intensity decrement due to the SZ effect is reported on the right panel of Figure \ref{sz images}. Each point shows the mean surface brightness in the concentric annuli of Figure \ref{sz images}, with uncertainties given by the standard deviation divided by the square root of the number of beams in each annulus. 
We estimated the S/N of the SZ decrement by defining the signal, $\Delta I_i$, as the difference between every decrement measurement and the outer plateau reached beyond $\sim1200$ kpc, of $(4.0 \pm 0.9)\times10^{-5}$ Jy beam$^{-1}$, then computing the noise as the quadrature sum of the uncertainties of the annuli used in the measurement,
\begin{equation}
    S/N=\frac{\sum\Delta I_i}{\sqrt{\sum \sigma_i^2}}
.\end{equation}
Thus, we obtained a S/N of 3. The cluster centre was defined by the peak SZ decrement. While SZ and X-ray centres can differ in merging systems, our SZ centre ($RA_{J2000}=17^{\rm h}52^{\rm m}0.5^{\rm s}$, $DEC_{J2000}=+44^\circ40'46''$) coincides, at our pixel scale, with the RASS position \citep[Table 1 in][]{bonafede2012}.

\section{Decrement modelling} \label{modeling}

The observed decrement was compared with a model prediction. At cm wavelengths, in the Rayleigh-Jeans
        approximation, \citet{basu2016} derived the theoretical decrement, $\Delta I_{SZ}$,

\begin{equation}
    \left( \frac{\Delta I_{SZ}}{\mathrm{mJy/beam}} \right)
    = \frac{1}{340} 
    \left( \frac{\nu}{\mathrm{GHz}} \right)^2
    \left( \frac{\Delta T_{\mathrm{RJ}}}{\mathrm{mK}} \right)
    \left( \frac{\Omega_{\mathrm{beam}}}{\mathrm{arcmin}^2} \right), \label{eq decrement} 
\end{equation}

\noindent
where $\nu$ is the observing frequency in GHz, $\Omega_{\mathrm{beam}}$ is the beam solid angle in square arcminutes, and $\Delta T_{RJ}$=-2$\cdot y \cdot T_{CMB}$ with $T_{CMB}$ the CMB temperature. The Compton y-parameter is defined as \citep[e.g.][]{birkinshaw1999}:

\begin{equation}
y = \int n_e(r) \sigma_T \frac{k_{\mathrm{B}} T_e(r)}{m_e c^2} \, dl ,
\end{equation}

\noindent
where $n_e$ is the electron density, $\sigma_T$ the Thomson scattering cross section, $T_e$ the ICM temperature, $m_e$ the electron mass, and $c$ the speed of light. We integrated along the line of sight, $l$.
The electron density has been assumed to follow a simple spherically symmetric $\beta$-model, described by

\begin{equation}
    n_e(r) = n_{0} \left(1 + \frac{r^2}{r_c^2} \right)^{-3\beta/2} , \label{electrondensity} 
\end{equation}
where $n_0$ is the central electron density of the cluster and $r_c$ is the core radius.\\
\noindent
A best-fit set of parameters $n_0$, $r_c$, and $\beta$ was obtained using a Markov Chain Monte Carlo (MCMC) approach with the $emcee$ sampler \citep{emcee} and uniform priors, assuming an isothermal model for the cluster with $k_B T_e=5.9$ keV \citep{finner2021}. The results are listed in Table~\ref{tab:retrieved_params}, with posterior distributions shown in the corner plot in Figure~\ref{corner}. The corner plot highlights the correlation between the $\beta$ parameter and the core radius, as we would expect from the model adopted. Nevertheless, the derived parameters are consistent with expectations for a cluster of the mass of MACS~J1752+4440 \citep[for studies on other clusters, see e.g. ][]{2002A&A...387...56P,bonamente2006}, demonstrating that even this simplified model can effectively offer a starting point in the characterisation of galaxy clusters.

\begin{table}
{\small

\captionsetup{font=footnotesize, skip=0pt}
\captionof{table}{Retrieved $\beta$-model parameters from MCMC sampling.}
\label{tab:retrieved_params}
\vspace{0.1cm}
\centering
\begin{tabular}{lccc}
\hline\hline 
    Parameter & Median & \textbf{$-\sigma$} & \textbf{$\sigma$} \\
    \hline
    $r_c$ [kpc] & 160 & $-30$ & $+30$ \\
    log $n_0$ [cm$^{-3}$] & -2.6 & $-0.1$ & $+0.1$ \\
    $\beta$ & 0.6 & $-0.1$ & $+0.1$ \\
    \hline
    \end{tabular}
    \tablefoot{Quoted values are medians with 16th and 84th percentile uncertainties.}}
\end{table}

\section{Discussion and conclusions} \label{conclusions}
We report the first SRT detection of the tSZ effect at 18.6 GHz in MACS J1752+4440, revealing a central surface-brightness decrement. The signal was modelled using the profile of \citet{basu2016}, adopting a spherical $\beta$ model for the electron density. An MCMC analysis was used to constrain the central density, core radius, and $\beta$ value. Given the merging nature of MACS J1752, these parameters should be considered as effective quantities, describing the average SZ signal, rather than the true 3D ICM structure; nevertheless, they are physically consistent with expectations for a cluster of this mass. To our knowledge, no estimates of these parameters have been published from X-ray data to date. By inverting Eq. \ref{eq decrement} and averaging within a 3.5$\arcmin$ radius (half the Planck beam), we obtained $y_{\rm SRT}^{0.9'}=(2.3 \pm 0.3)\cdot10^{-5}$. Convolving the SRT map with the 7$\arcmin$ Planck beam yields $y_{\rm SRT}^{7\arcmin}=(1.8 \pm 0.3)\cdot10^{-5}$, higher than the Planck MILCA value of $y_{\rm Planck}=(4\pm1)\cdot10^{-6}$ \citep{planck2016}.
 \footnote{Data taken from MILCA \citep{hurier2013} 7\arcmin map, \url{https://szdb.osups.universite-paris-saclay.fr/ymap.html}.} This discrepancy is largely explained by angular-resolution effects. Planck’s 7$\arcmin$ beam smooths the signal on cluster scales, leading to an underestimation of Compton-$y$ by factors of $\sim$3 \citep[e.g.][]{pact2019A&A...632A..47A}. The higher resolution of the SRT also enables subtraction of contaminating radio sources: a typical $\sim$1 mJy source contributes $\Delta y\sim10^{-6}$ (Eq.~\ref{eq decrement}, Planck beam at $\sim$100 GHz), comparable to Planck sensitivity, but negligible after a correction on the SRT map. Our simplified spherical $\beta$ model introduces additional effects. Neglecting cluster geometry or any departures from isothermality can bias SZ- and X-ray-derived parameters by up to $\sim 30\%$ \citep{Puy2000}. Assuming an infinite line-of-sight depth further overestimates $y_{\rm SRT}^{7\arcmin}$ by projecting more gas than is physically present. Departures from spherical symmetry can modify both the amplitude and the shape of the inferred Compton-$y$ profile: an elongation along the line of sight enhances the SZ signal, while asymmetries in the plane of the sky are smoothed by azimuthal averaging, producing a flatter effective profile. Such considerations underscore the need for cluster-specific modelling, while supporting the robustness of the SRT detection given the model simplifications and the different angular resolution of Planck. Previous SZ measurements at 15–20 GHz have mainly relied on interferometric observations \citep[e.g.][]{massardi2010ApJ...718L..23M,rodriguez2011MNRAS.414.3751A}, which are intrinsically limited in their sensitivity to diffuse emission on arcminute scales due to missing short baselines. The SRT K-band observations probe a complementary regime, providing sensitivity to extended SZ signals on arcminute scales at low frequencies and enabling the joint characterisation of thermal and non-thermal ICM components on cluster scales, which is a frequency-angular-scale parameter space that is not fully covered by current higher resolution interferometric or mm-wave facilities. At 18.6 GHz, these observations only probe  the low-frequency tail of the SZ spectral distortion, which reaches its minimum near 150 GHz and, therefore, they are not intended as primary means for SZ studies. However, they demonstrate the high sensitivity of the SRT and its capacity to detect and characterise SZ signals at low frequencies at a higher resolution than Planck. Moreover, the detection of the SZ decrement during observations aimed at studying diffuse cluster emission represents an added value of such datasets: a preliminary estimate of the Compton-$y$ parameter can inform the selection of promising targets for follow-up observations; for instance, at 90~GHz with the MISTRAL receiver at the SRT.

\begin{acknowledgements}
The Enhancement of the Sardinia Radio Telescope (SRT) for the study of the Universe at high radio frequencies is financially supported by the National Operative Program (Programma Operativo Nazionale - PON) of the Italian Ministry of University and Research “Research and Innovation 2014-2020”, Notice D.D. 424 of 28/02/2018 for the granting
of funding aimed at strengthening research infrastructures,
in implementation of the Action II.1 – Project Proposals
PIR01 00010 and CIR01 00010. This work was carried out thanks to the funding of the Regione Autonoma della Sardegna, ai sensi della Legge Regionale 7 agosto 2007, n.7
"Promozione della Ricerca Scientifica e dell'Innovazione Tecnologica in Sardegna. SC acknowledges support from the ERC CoG $\vec{B}$ELOVED, GA N.101169773. We thank the referee for comments and suggestions.
\end{acknowledgements}
\bibliographystyle{aa}   
\bibliography{vfc}

\clearpage

\begin{appendix}

\onecolumn
\section{Parameter retrieval: Corner plot}

\begin{figure*}[h!]
    \centering
     \resizebox{12cm}{13cm}
    {\includegraphics {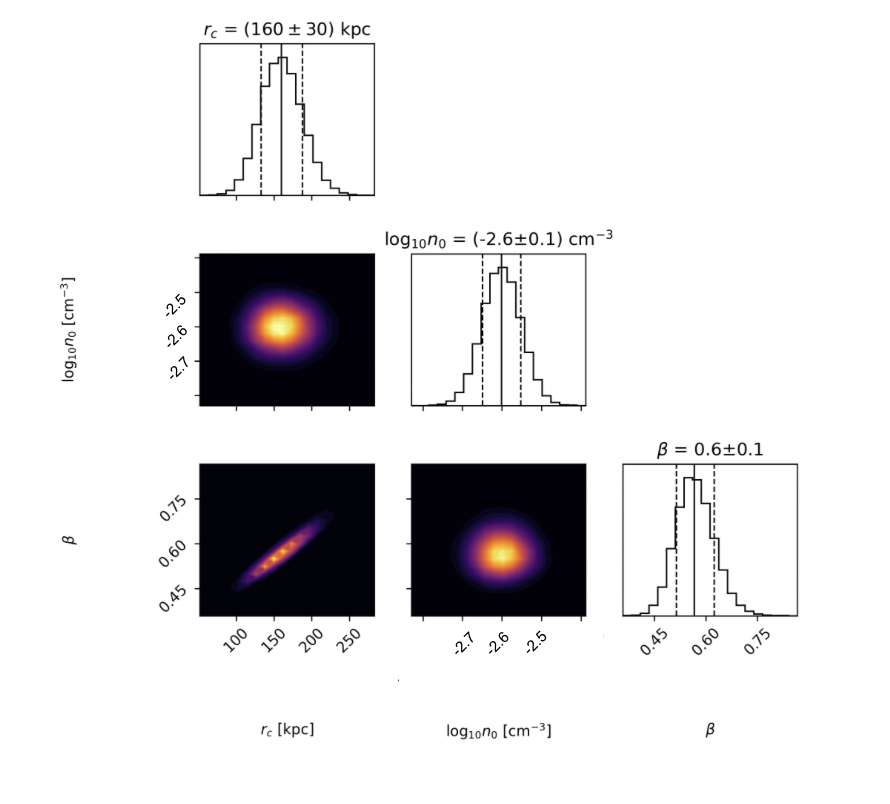}}
     \caption{Retrieval of parameters for the density profile using $emcee$. The continuous line represents the median of the distribution. The dashed lines show the 16th and 84th percentiles.}
      \label{corner}
\end{figure*}

\end{appendix}
\end{document}